\documentstyle[12pt]{article}

\catcode`\@=11
\long\def\@makefntext#1{ 
\protect\noindent \hbox to 3.2pt {\hskip-.9pt
$^{{\ninerm\@thefnmark}}$\hfil}#1\hfill} 

\def\thefootnote{\fnsymbol{footnote}}
 \def\@makefnmark{\hbox to 0pt{$^{\@thefnmark}$\hss}}  

\def\ps@myheadings{\let\@mkboth\@gobbletwo
\def\@oddhead{\hbox{} 
\rightmark\hfil\ninerm\thepage}
\def\@oddfoot{}\def\@evenhead{\ninerm\thepage\hfil 
\leftmark\hbox{}}\def\@evenfoot{}
\def\sectionmark##1{}\def\subsectionmark##1{}}

\begin{document}

\newcommand{\symbolfootnote}{\renewcommand{\thefootnote}
	{\fnsymbol{footnote}}}
\renewcommand{\thefootnote}{\fnsymbol{footnote}}
\newcommand{\alphfootnote}
	{\setcounter{footnote}{0}
	 \renewcommand{\thefootnote}{\sevenrm\alph{footnote}}}

\newcounter{sectionc}\newcounter{subsectionc}\newcounter{subsubsectionc}
\renewcommand{\section}[1] {\vspace{0.6cm}\addtocounter{sectionc}{1}
\setcounter{subsectionc}{0}\setcounter{subsubsectionc}{0}\noindent
	{\bf\thesectionc. #1}\par\vspace{0.4cm}}
\renewcommand{\subsection}[1] {\vspace{0.6cm}\addtocounter{subsectionc}{1}
	\setcounter{subsubsectionc}{0}\noindent
	{\it\thesectionc.\thesubsectionc. #1}\par\vspace{0.4cm}}
\renewcommand{\subsubsection}[1] {\vspace{0.6cm}
\addtocounter{subsubsectionc}{1}
	\noindent {\rm\thesectionc.\thesubsectionc.\thesubsubsectionc.
	#1}\par\vspace{0.4cm}}
\newcommand{\nonumsection}[1] {\vspace{0.6cm}\noindent{\bf #1}
	\par\vspace{0.4cm}}

\newcounter{appendixc}
\newcounter{subappendixc}[appendixc]
\newcounter{subsubappendixc}[subappendixc]
\renewcommand{\thesubappendixc}{\Alph{appendixc}.\arabic{subappendixc}}
\renewcommand{\thesubsubappendixc}
	{\Alph{appendixc}.\arabic{subappendixc}.\arabic{subsubappendixc}}

\renewcommand{\appendix}[1] {\vspace{0.6cm}
        \refstepcounter{appendixc}
        \setcounter{figure}{0}
        \setcounter{table}{0}
        \setcounter{equation}{0}
        \renewcommand{\thefigure}{\Alph{appendixc}.\arabic{figure}}
        \renewcommand{\thetable}{\Alph{appendixc}.\arabic{table}}
        \renewcommand{\theappendixc}{\Alph{appendixc}}
        \renewcommand{\theequation}{\Alph{appendixc}.\arabic{equation}}
        \noindent{\bf Appendix \theappendixc #1}\par\vspace{0.4cm}}
\newcommand{\subappendix}[1] {\vspace{0.6cm}
        \refstepcounter{subappendixc}
        \noindent{\bf Appendix \thesubappendixc. #1}\par\vspace{0.4cm}}
\newcommand{\subsubappendix}[1] {\vspace{0.6cm}
        \refstepcounter{subsubappendixc}
        \noindent{\it Appendix \thesubsubappendixc. #1}
	\par\vspace{0.4cm}}

\def\abstracts#1{{
	\centering{\begin{minipage}{30pc}\tenrm\baselineskip=12pt\noindent
	\centerline{\tenrm ABSTRACT}\vspace{0.3cm}
	\parindent=0pt #1
	\end{minipage} }\par}}

\newcommand{\bibit}{\it}
\newcommand{\bibbf}{\bf}
\renewenvironment{thebibliography}[1]
	{\begin{list}{\arabic{enumi}.}
	{\usecounter{enumi}\setlength{\parsep}{0pt}
\setlength{\leftmargin 1.25cm}{\rightmargin 0pt}
	 \setlength{\itemsep}{0pt} \settowidth
	{\labelwidth}{#1.}\sloppy}}{\end{list}}

\topsep=0in\parsep=0in\itemsep=0in
\parindent=1.5pc

\newcounter{itemlistc}
\newcounter{romanlistc}
\newcounter{alphlistc}
\newcounter{arabiclistc}
\newenvironment{itemlist}
    	{\setcounter{itemlistc}{0}
	 \begin{list}{$\bullet$}
	{\usecounter{itemlistc}
	 \setlength{\parsep}{0pt}
	 \setlength{\itemsep}{0pt}}}{\end{list}}

\newenvironment{romanlist}
	{\setcounter{romanlistc}{0}
	 \begin{list}{$($\roman{romanlistc}$)$}
	{\usecounter{romanlistc}
	 \setlength{\parsep}{0pt}
	 \setlength{\itemsep}{0pt}}}{\end{list}}

\newenvironment{alphlist}
	{\setcounter{alphlistc}{0}
	 \begin{list}{$($\alph{alphlistc}$)$}
	{\usecounter{alphlistc}
	 \setlength{\parsep}{0pt}
	 \setlength{\itemsep}{0pt}}}{\end{list}}

\newenvironment{arabiclist}
	{\setcounter{arabiclistc}{0}
	 \begin{list}{\arabic{arabiclistc}}
	{\usecounter{arabiclistc}
	 \setlength{\parsep}{0pt}
	 \setlength{\itemsep}{0pt}}}{\end{list}}

\newcommand{\fcaption}[1]{
        \refstepcounter{figure}
        \setbox\@tempboxa = \hbox{\tenrm Fig.~\thefigure. #1}
        \ifdim \wd\@tempboxa > 6in
           {\begin{center}
        \parbox{6in}{\tenrm\baselineskip=12pt Fig.~\thefigure. #1 }
            \end{center}}
        \else
             {\begin{center}
             {\tenrm Fig.~\thefigure. #1}
              \end{center}}
        \fi}

\newcommand{\tcaption}[1]{
        \refstepcounter{table}
        \setbox\@tempboxa = \hbox{\tenrm Table~\thetable. #1}
        \ifdim \wd\@tempboxa > 6in
           {\begin{center}
        \parbox{6in}{\tenrm\baselineskip=12pt Table~\thetable. #1 }
            \end{center}}
        \else
             {\begin{center}
             {\tenrm Table~\thetable. #1}
              \end{center}}
        \fi}

\def\@citex[#1]#2{\if@filesw\immediate\write\@auxout
	{\string\citation{#2}}\fi
\def\@citea{}\@cite{\@for\@citeb:=#2\do
	{\@citea\def\@citea{,}\@ifundefined
	{b@\@citeb}{{\bf ?}\@warning
	{Citation `\@citeb' on page \thepage \space undefined}}
	{\csname b@\@citeb\endcsname}}}{#1}}

\newif\if@cghi
\def\cite{\@cghitrue\@ifnextchar [{\@tempswatrue
	\@citex}{\@tempswafalse\@citex[]}}
\def\citelow{\@cghifalse\@ifnextchar [{\@tempswatrue
	\@citex}{\@tempswafalse\@citex[]}}
\def\@cite#1#2{{$\null^{#1}$\if@tempswa\typeout
	{IJCGA warning: optional citation argument
	ignored: `#2'} \fi}}
\newcommand{\citeup}{\cite}

\def\fnm#1{$^{\mbox{\scriptsize #1}}$}
\def\fnt#1#2{\footnotetext{\kern-.3em
	{$^{\mbox{\sevenrm #1}}$}{#2}}}

\font\twelvebf=cmbx10 scaled\magstep 1
\font\twelverm=cmr10 scaled\magstep 1
\font\twelveit=cmti10 scaled\magstep 1
\font\elevenbfit=cmbxti10 scaled\magstephalf
\font\elevenbf=cmbx10 scaled\magstephalf
\font\elevenrm=cmr10 scaled\magstephalf
\font\elevenit=cmti10 scaled\magstephalf
\font\bfit=cmbxti10
\font\tenbf=cmbx10
\font\tenrm=cmr10
\font\tenit=cmti10
\font\ninebf=cmbx9
\font\ninerm=cmr9
\font\nineit=cmti9
\font\eightbf=cmbx8
\font\eightrm=cmr8
\font\eightit=cmti8

       {\normalsize \hfill
       \begin{tabbing}
       \`\begin{tabular}{l}
         hep--th/9408137 \\
         SLAC--PUB--66--40  \\
         August 1994 \\
         (T) \\
         \end{tabular}       
       \end{tabbing} }\vspace{8mm}
\thispagestyle{empty}

\centerline{\twelvebf Quantum Corrections for a Cosmological String 
 Solution\footnote{Work supported by the Department of Energy, 
 contract DE--AC03--76SF00515.}}
\vspace{0.8cm}
\centerline{\tenrm Klaus Behrndt\footnote{E-Mail: 
                           behrndt@jupiter.slac.stanford.edu}}
\baselineskip=13pt
\centerline{\tenit Stanford Linear Accelerator Center}
\baselineskip=12pt
\centerline{\tenit Stanford University, Stanford, California 94309, USA}
\vspace{0.9cm}

\renewcommand{\arraystretch}{2.0}
\renewcommand{\thefootnote}{\alph{footnote}}
\newcommand{\be}{\begin{equation}}
\newcommand{\ee}{\end{equation}}
\newcommand{\ba}{\begin{array}}
\newcommand{\ea}{\end{array}}
\newcommand{\vsf}{\vspace{5mm}}
\newcommand{\NP}[3]{{\em Nucl. Phys.}{ \bf B#1#2#3}}
\newcommand{\PRD}[2]{{\em Phys. Rev.}{ \bf D#1#2}}
\newcommand{\MPLA}[1]{{\em Mod. Phys. Lett.}{ \bf A#1}}
\newcommand{\PL}[3]{{\em Phys. Lett.}{ \bf B#1#2#3}}
\newcommand{\marpar}{\marginpar[!!!]{!!!}}
\newcommand{\lab}[2]{\label{#1#2}   (#1#2) \hfill }
\input epsf.tex
\vspace{10mm}

\abstracts{
We investigate quantum corrections for a cosmological solution 
of the string effective action. Starting point is a classical solution 
containing an antisymmetric tensor field, a dilaton and a modulus
field which has singularities in the scalar fields. As a first step
we quantize the scalar fields near the singularity with the result
that the singularities disappear and that in general non-perturbative
quantum corrections form a potential in the scalar fields. 
}
\vspace{10mm}

\vfill
\baselineskip=14pt

\begin{center}
Invited talk presented at \\
{\em Particles, Strings and Cosmology (PASCOS)} \\
Syracuse, NY, May 19 -- 24, 1994
\end{center}

\vfill

\newpage

\twelverm   
\baselineskip=14pt
\section{Introduction}
\vspace*{-0.35cm}
In two dimension (2-D) the dilaton gravity could be formulated as a
quantum theory in the last years.\cite{cghs,bu/ch,strom,rst,deal} This
opens the possibility to quantize higher dimensional theories near
regions where a 2-D part factorizes. As a first step one can quantize
only the 2-D part and leave the dynamical fields living in the other
dimensions as a classical background. This procedure is especially
motivated if the 2-D part contains singularities and therefore quantum
corrections are expected to become important whereas the other part
behaves smooth. In this paper we are going to describe one example
with this property. It is a solution of the 5-D string effective
action
\be   \label{01}  
S^{(5)} = \int d^5 x \sqrt{G} e^{-2 \psi} \left( R + 4 (\partial \psi)^2
  - \frac{1}{12} H^2 \right)
\ee
where $\psi$ is the dilaton field and $H_{\mu\nu\lambda}  =
\partial_{[\mu} B_{\nu\lambda]}$ is the torsion corresponding to the
antisymmetric tensor field $B_{\mu\nu}$. The 5-D metric in our example
has the following form
\be  \label{02}
ds^2 = \rho^2 dw^2 - e^{\lambda} dt^2 + t^2 d\Omega_{k}^2 =
 \rho^2 dw^2 + \tilde{G}_{\mu\nu} dx^{\mu} dx^{\nu}
\ee
where $d\Omega_{k}^2$ is the volume measure corresponding to a 3-D
space with constant curvature $k$ (1, 0, -1).  In section 2 we discuss
the (classical) solution in detail. It turns out that the functions
$\rho$ and $\lambda$ depend only on $t$ and that the dilaton and torsion
are independent of the fifth coordinate $y$.  So, it is
possible to reduce the 5-D theory (\ref{01}) down to a 4-D string
theory
\be  \label{03}
S^{(5)} \rightarrow S^{(4)} = \int d^4x \sqrt{\tilde{G}} e^{-2 \phi}
 \left( \tilde{R} + 4 (\partial\phi)^2 -
(\frac{\partial \rho}{\rho})^2 - \frac{1}{12} H^2 \right) 
\ee
where $\phi=\psi -\frac{1}{2} \log \rho $ is the 4-D dilaton field and
$\rho$ is the modulus field. After this Kaluza--Klein reduction we end
up with a 4-D Friedman--Robertson--Walker (FRW) metric with spatial
curvature $k$. This 5-D Kaluza--Klein approach in the string theory
corresponds to a compactification where not all other
dimensions\footnote{We are working in a critical string theory,
i.e.~in the critical space time dimension (e.g.~26 for the bosonic
case).} \ are compactified on a torus with constant radii. Instead,
one of these interior dimensions has a non-constant compactification
radius $\rho$.

In the literature there are some cosmological Kaluza-Klein solutions
discussed.  Similar to this approach Matzner, Mezzacappa and
Wiltshire\cite{mmw} have used the way through a 5-D theory to find new
classical solutions for the Einstein gravity in 4-D.  For the string
effective action Copeland, Lahiri and Wands\cite{clw} were able to
find a solution even for higher dimensional interior
space. The qualitative behavior is similar to the classical solution
we are going to describe in the next section.  In the third section we
are going to discuss the quantized theory after a s-wave reduction
near the singularity and, finally, we summarize our results.

\newpage
\section{Classical Theory}
The classical solution we are interested in is described in Ref.~8.
Let us summarize the main features in this section.  The original
motivation was to discuss the FRW cosmology in a string Kaluza-Klein
approach. If one wants to do this one should first discuss the
symmetries. The FRW metrics are the most general metrics describing a
spatial homogeneous and isotropic universe. The geometry of the 3-D
spatial part is therefore given by a $S_{k}^3$ which is for $k=+1$ a
sphere, for $k=-1$ a pseudo-sphere or for $k=0$ a flat space. If we now
embed this 4-D space in a higher dimensional space time we have to
make sure that the higher dimensional metric respects this
symmetry. In this case, it means that the 5-D metric has to have the
$S_{k}^3$ spherical symmetry and thus can be written as a
Schwarzschild metric, i.e.~the 5-D metric has to have the form
(\ref{02}). Furthermore, because we are interested in a 5-D theory
which can be reduced to a 4-D FRW solution we have to look for
solutions which are independent of the fifth coordinate $w$. If we
start with the spherical case ($k=+1$) and remember that the time in
Eq.~(\ref{02}) is the radius of the $S^3$ we have simply to look for a
static black hole solution. Then we have to replace the time in the
black hole solution by our fifth coordinate, the radius by our time
and we have to switch the signature in the metric correspondingly. By
that way we have a solution for $k=+1$ which we have finally to
generalize to arbitrary $k$. Following this procedure and starting
with the 5-D black hole solution discussed by Strominger and
Horowitz\cite{st/ho}\ \footnote{This solution was first found by
Gibbons and Maeda\cite{gi/ma} \ for the higher dimensional
case.} \ we find as general solution
\be  \label{04}
\ba{c}
ds^2 = \frac{-k+\left(\frac{t_+}{t}\right)^2}{1-\left(\frac{t_-}{t}\right)^2}
     dw^2 - \frac{1}{(-k+(\frac{t_+}{t})^2)(1-(\frac{t_-}{t})^2)} dt^2
    + t^2 d\Omega_{k}^2 \\
e^{-2(\psi - \psi_{0})} = 1 - \left(\frac{t_-}{t}\right)^2 \qquad , \qquad
  H = 2 t_+ t_- \epsilon_{3,k} \ea \ee 
where the prefactor in the torsion $2 t_- t_+ =Q_M$ defines a magnetic
charge and $\epsilon_{3,k}$ is the volume form corresponding to
$d\Omega_k$, i.e.\ 
$$ \epsilon_{3,k}=\left( \frac{\sin \sqrt{k}
\chi}{\sqrt{k}}\right)^2 \sin \theta \, d\chi \wedge d\theta \wedge
d\varphi \ .$$
After the dimensional reduction we find for the 4-D metric and dilaton
\begin{equation} \label{05}
\begin{array}{l c l}
\tilde{ds}^2  =  -\frac{dt^2}{\left( -k+\left(\frac{t_+}{t}\right)^2\right)
\left( 1-\left(\frac{t_-}{t}\right)^2\right)}+t^2d\Omega_k^2 & , &
e^{-2\phi}  \sim  \sqrt{ \left( 1-\left(\frac{t_-}{t}\right)^2\right)
\left( -k+\left(\frac{t_+}{t}\right)^2\right)} \ .
\end{array} \end{equation}
Obviously, this solution is not well defined for arbitrary $t$. In order
to ensure a real dilaton and the right signature in the metric
we have instead to restrict the allowed $t$ region to
\begin{equation}   \label{06}
\left( -k +\left(\frac{t_+}{t}\right)^2\right)
\left(1 -\left(\frac{t_-}{t}\right)^2\right) > 0.
\end{equation}
However, this restriction does not mean that the lifetime is finite,
but since $t$ defines the radius of the $S_{k}^3$, it means that the
spatial extension is bounded. Especially, we have a lower bound for 
all $k$ as long as $t_-$ is non-vanishing, i.e.,as long as
the magnetic charge is non-vanishing (let us assume that, without any
restriction, $t_- < t_+$). Since for FRW metrics singularities
correspond to zeros or singularities in the world radius we see, that
a non-vanishing magnetic charge ($t_- \neq 0$) prevents the universe
from forming a singularity. This becomes clear if we transform the
metric in the conformal time
\be \label{07}
\tilde{ds}^2 = \left\{ t_{-}^2 + (t_{+}^2 - k t_{-}^2) 
 \left(\frac{\sin \sqrt{k}
  \eta }{\sqrt{k}}\right)^2 \right\} \, \left[ -d\eta^2 +
  d\Omega_{k}^2 \right] \ .
\ee
This solution oscillates for $k=+1$ between the minimum
$t_-$ and the maximum $t_+$, for $k=-1$ this solution is asymptotically
($\eta \rightarrow \infty$) flat and has wormhole at $\eta =0$.
For $k=0$ we have no flat regions, but again, the solution shrinks
until it reaches at $\eta = 0$ the non-vanishing minimum in order
to expand then again. Unfortunately, it is not possible to find
an analytic expression for the world radius $a(\tau)$ in the
standard parameterization of the FRW metric $ds^2 =- d\tau^2 +
a^2(\tau) d\Omega_{k}^2$. We have plotted numerical results in
figure 1.

\begin{figure}[t] \vspace*{-3mm}
\hspace{-5mm}
\begin{tabular}{lr}
 \begin{minipage}[t]{75mm}
 \epsfxsize=70mm
  \leavevmode 
\epsffile{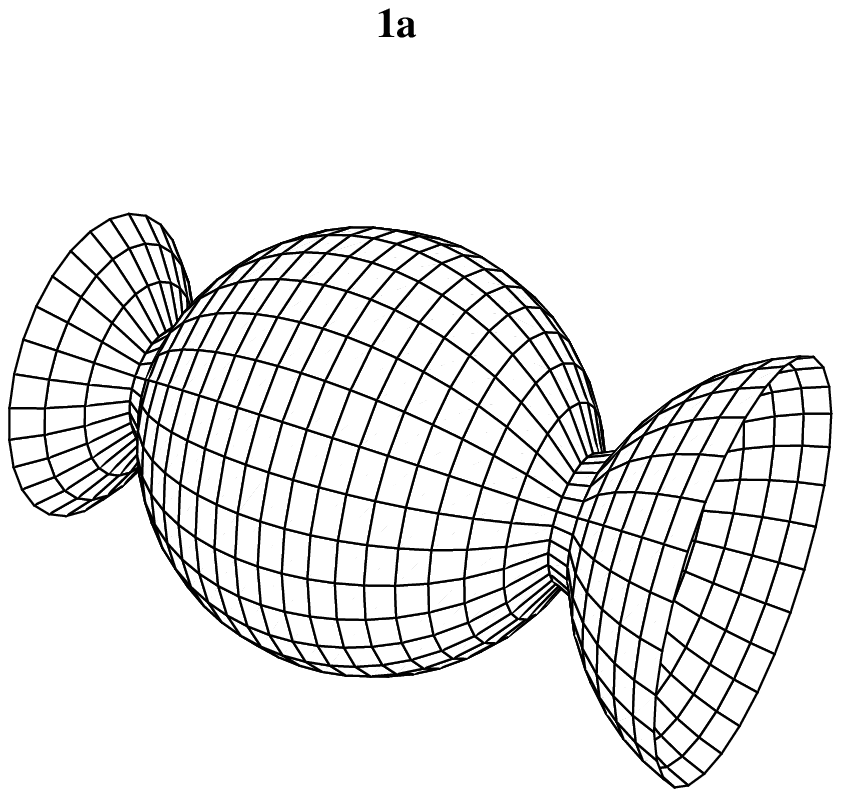}
     \end{minipage}
 \hspace{10mm} &
  \begin{minipage}[t]{75mm}
\epsfxsize=70mm
  \leavevmode 
     \epsffile{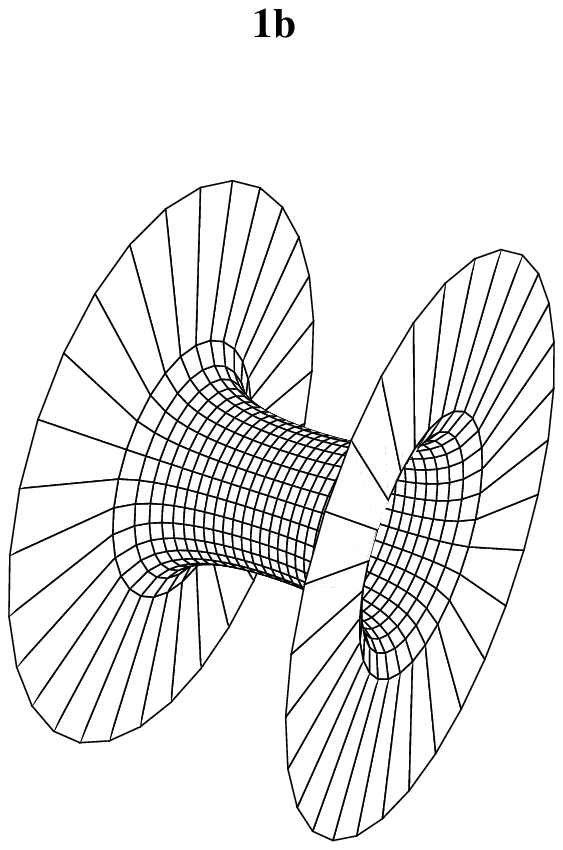}
 \end{minipage}
\end{tabular} \vspace{-10mm}
\caption{In (a) we have plotted the closed oscillating solution
for $k=1$;  (b) is the wormhole solution for $k=-1$.}
\vspace{2mm}
\end{figure}

Remarkably, although the 4-D metric is completely smooth the scalar
fields contain divergencies. As one can see in Eq.~(\ref{04}) the modulus
field $\rho$ is infinite at $t^2=t_{-}^2$ (or $\eta =0$) and shrinks
then with the time. This means, at the minimal extension of the world
radius, e.g.~inside the wormhole, the compactification radius $\rho$
is infinite, i.e.~there is no compactification. With the further time
evolution the world radius expands and the compactification radius
shrinks, which means that we have a dynamical compactification. This
behavior seems to be quite general. The qualitative feature remains
intact also for more than one additional dimensions\cite{clw}.
Similarly, the dilaton goes to $+\infty$ at the extrema of the world
radius $t^2 = t_{\pm}^2$. There, the theory is in the strong coupling
region. The consequence of this behavior is that the Einstein metric
becomes singular at these points.  This metric for which the effective
action has the Einstein-Hilbert term as the first part is defined by
the field redefinition
\begin{equation} \label{08}
G_{\mu \nu}^{(E)} = e^{-2\phi}\, \tilde{G}_{\mu \nu} \ .
\end{equation}
The vanishing Weyl factor is responsible for the singularity. 
For a better understanding of this singularities let us discuss
the 5-D metric. In the conformal time this metric is given by
\be \label{09}
\ba{l}
ds^2 = \left( \frac{\sqrt{k}}{\tan \sqrt{k} \eta } \right)^2 \, dw^2
  + \left\{ t_{-}^2 + (t_{+}^2 - k t_{-}^2) \left(\frac{\sin \sqrt{k}
  \eta }{\sqrt{k}}\right)^2 \right\} \, \left[ -d\eta^2 +
  d\Omega_{k}^2 \right] \\
 e^{2 (\psi - \psi_0)} = 1 + \frac{t_{-}^2}{(t_{+}^2 - k t_{-}^2)
 \left(\frac{\sin \sqrt{k} \eta }{\sqrt{k}}\right)^2}\ .
\ea
\ee
If we now approach the singularity at ($\sin \sqrt{k} \eta \simeq 0$)
we find that the 5-D metric factorizes in a 2-D ($w,\eta$) part
and a 3-D spherical part
\be  \label{10}
ds^2 = \left( \frac{\sqrt{k}}{\tan \sqrt{k} \eta } \right)^2 \, dw^2
  - t_{-}^2 d\eta^2 \ + \  t_{-}^2 \, d\Omega_{k}^2  \quad , \quad
 e^{2 \psi} \sim \frac{1}
{\left(\frac{\sin \sqrt{k} \eta }{\sqrt{k}}\right)^2} \ .
\ee
We see, that the singularities are contained in the 2-D part, which
is just the known black hole solution\cite{witt} and that the 3-D
spherical part behaves smooth. Before we turn to quantize this
theory let us make one remark here. One can ask whether there
is a limit in which the theory goes over in a conformal field
theory (CFT). In a  previous paper\cite{behr} we have shown that
this limit is given by $t_{+}^2\rightarrow k t_{-}^2$ and a suitable
constant shift in the dilaton. Then the 5-D solution becomes (\ref{10}) 
for all $\eta$. The result is the direct product of the
CFT which is behind the black solution (SL(2,R)/U(1) WZW model) and
a 3-D parallized space (for $k=+1$ it is the SU(2) WZW model).
Because the spherical part does not depend on the time, in this
extremal limit the solution becomes static  in the string metric.
But, nevertheless, the Einstein metric receives a time dependence
via the non-constant dilaton.

\section{Quantization and Dilaton/Moduli Potential}
In this section we investigate the quantum theory near the
singularity. In the figure it is just the region of minimal extension,
e.g.~inside the wormhole of figure 1b. As we pointed out in the
introduction we are not going to quantize this theory
completely. Instead, as a first step, we quantize the singular 2-D part
and leave the spherical part as a classical background. During this
procedure, which is also known as s-wave reduction, we integrate out
the spherical degrees of freedom. This is motivated by the assumption
that the quantum corrections respect the spherical symmetry. Generally,
this is not the case, but practicable as a first approximation. 
A further motivation comes from the 4-D theory. The quantization
of the 2-D metric and dilaton only corresponds there to a quantization
of the scalar fields (moduli and dilaton).  This means,  our approach 
corresponds to the quasi-classical approximation 
\be \label{11}
R_{\mu\nu}^{(E)} - \frac{1}{2} R^{(E)} G_{\mu\nu}^{(E)} = <T_{\mu\nu}^{(\phi,
 \rho)}> + T_{\mu\nu}^{(H)}
\ee
where $G_{\mu\nu}^{(E)}$ is the metric in the Einstein frame (\ref{08}).
When quantizing this theory we are especially interested
in what happens with the singularity and whether quantum corrections
can form a dilaton/moduli potential. 

After the s-wave reduction we get 
\be \label{12}
S^{(5)} \rightarrow S^{(2)} = \int d^2z \sqrt{g} e^{-2 \phi} \left( R^{(2)} 
+ 4 (\partial  \phi)^2 + \lambda \right) \ .
\ee
This is the known dilaton gravity with $\lambda = \frac{2}{t_-^2} 
\left(3k - (\frac{t_+}{t_-})^2 \right)$.  As a consistency condition we
have to ensure that in the classical limit (weak coupling limit, $\phi
\rightarrow - \infty$) we get the back the classical solution, which
is in conformal coordinates given by \cite{witt}
\be  \label{13}
ds^2 = e^{2 \sigma} dz^+ dz^-  \qquad , \qquad e^{-2 \phi} \sim e^{-2 \sigma}
= u - \lambda z^+ z^-
\ee
where $u$ is constant. This solution can be transformed in the 2d
($w,\eta$) part of Eq.~(\ref{10}) where $\eta \simeq 0$ corresponds to $u
\simeq \lambda z^+ z^-$. 

We are now following the procedure of de Alwis\cite{deal}. 
Choosing the conformal gauge
\be  \label{14}  
g_{ab} = e^{2 \sigma} \hat{g}_{ab}
\ee
we can write (\ref{12}) as a general 2d $\sigma$ model
\be \label{15}
S^{(2)} = -\int d^2 z \sqrt{\hat{g}} \left[ \hat{g}^{ab}\partial_a X^{\mu}
  \partial_b X^{\nu} G_{\mu\nu}(X) + \hat{R}\, \Phi(X) + T(X) \right]
\ee
with: $X^{\mu}=\{\phi,\sigma\}$. Thus, the quantization of the dilaton
gravity is reduced to the quantization of a 2d $\sigma$ model with the
target space spanned by $\phi$ and $\sigma$. 
We can write the metric $G_{\mu\nu}$ as
\be  \label{16}
dS^2 = -4 e^{-2\phi} [1+h(\phi)] d\phi^2 + 4 e^{-2 \phi} [1+\bar{h}(\phi)]
d\sigma d\phi +  \kappa d\sigma^2 
\ee
where $h$ and $\hat{h}$ are model dependent functions of $\phi$ or
$X^1$, which contain the quantum corrections. For $h=\bar{h}=0$ we have
the CGHS model \cite{cghs}; for $ 2 h = \bar{h} = -e^{2\phi}$ the
model discussed by Strominger\cite{strom}; $h=0$ and $\bar{h} = -
\frac{\kappa} {4} e^{2 \phi}$ describes the RST model \cite{rst}. The
parameter $\kappa=\frac{24-N}{6}$ originates from the definition of
the functional integration measure and $N$ corresponds to additional
conformal matter. In order to get a flat metric $G_{\mu\nu}$ we
introduce as next step new target space coordinates
\be  \label{17}
\ba{l}
x = \frac{2}{\sqrt{\kappa}} \int d\phi e^{-2 \phi} \sqrt{(1+\bar{h})^2
  +\kappa e^{2 \phi} (1+h)} \\
y = \sqrt{\kappa} \sigma - \frac{1}{\kappa} e^{-2 \phi} + \frac{2}
{\kappa} \int d\phi e^{-2 \phi} \bar{h} \ .
\ea
\ee
After this we end up with the model
\be \label{18}
S^{(2)} = \int d^2 z \sqrt{\hat{g}} \left[ (\partial x)^2 - (\partial y)^2
 + \hat{R} \, \Phi(x,y) + T(x,y) \right] \ .
\ee
However, the function $\Phi(x,y)$ and $T(x,y)$ are not arbitrary.
The requirement of independence of the reference metric $\hat{g}_{ab}$
has the consequence that the 2-D theory has to be conformal invariant. 
The simplest choice is to take a linear dilaton $\Phi$ and a exponential
tachyon $T$
\be \label{19}
\Phi = \sqrt{\kappa} y \qquad , \qquad  
T = \lambda e^{\frac{2}{\sqrt{\kappa}} (x-y)} \ .
\ee
With this choice we have a well defined 2-D quantum theory
(mathematically the same as the non-critical string theory in one
dimension). Now, one defines the quantum theory in
terms of these $x$ and $y$ variables and  regards Eq.~(\ref{12}) as the
classical limit. In order to investigate the influence on the 
classical solution we have to discuss the equation of motion
for $x$ and $y$ ($\hat{R}^{(2)}=0$)
\be  \label{20}
  \partial^2 x = \lambda \frac{2}{\sqrt{\kappa}} e^{-\frac{2}
{\sqrt{\kappa}} (x - y)} \qquad , \qquad \partial^2 y = \partial^2 x \ .
\ee
Solving these equations we have to restrict ourselves to solutions
that reproduce in the classical limit the black hole solution (\ref{13}).
Therefore, we are interested in a solution depending on the product 
$z^+ z^-$ only and find
\be  \label{21}
x = y = \frac{1}{\sqrt{\kappa}} \left(u - \lambda z^+ z^- \right)
\ee
($u=const.$). Using the transformation (\ref{17}) we can express this
solution by $\phi$ and $\sigma$. In doing so we have to fix the up to
now arbitrary functions $h(\phi)$ and $\bar{h}(\phi)$.  Let us discuss
the parameterization suggested by de Alwis: $h=0$,
$\bar{h}=-\frac{1}{2} \kappa e^{2 \phi}$. This choice is motivated by
the fact that for all values of $\phi$ and $\sigma$ the transformation
(\ref{17}) is non-singular and secondly that the range of $x$ and $y$
goes from $-\infty$ to $+\infty$ if $\phi$ and $\sigma$ do so.  For
$x$ and $y$ one gets
\be  \label{22}
\ba{l}
x = \frac{1}{\sqrt{4 \kappa}}\left( - \sqrt{\kappa^2 + 4 e^{-4 \phi}}
  + \sqrt{\kappa} \mbox{arcsinh} \frac{\kappa}{2} e^{2\phi} \right) \\
y = \sqrt{\kappa} \left( \sigma - \frac{1}{\kappa} e^{-2 \phi}
  - \phi \right) 
\ea 
\ee 
In terms of Eqs.~(\ref{21}) one finds in the weak coupling limit ($e^{2
\phi} \ll 1$) the desired classical solution (\ref{13})
\be  \label{23}
e^{-2\phi} = u - \lambda z^+ z^- \qquad , \qquad \sigma = \phi\ .
\ee
But our original singularity appeared in the strong coupling region
(see Eq.~(\ref{10})). In this limit ($e^{2 \phi} \gg 1$) we obtain
\be  \label{24}
\phi = - \frac{1}{\kappa} ( u - \lambda z^+ z^-) \qquad , \qquad \sigma =
  \frac{1}{\kappa} e^{-2 \phi} \ .
\ee
Therefore, after incorporation of quantum corrections ($\sim
{\cal{O}}(e^{2 \phi})$) the black hole solution becomes smooth also in the
strong coupling region\cite{deal}. Note, that in the dilaton gravity a
singularity in the metric has to be accompanied by a singularity in
the dilaton, i.e.\ singularities can only appear in the strong or weak
coupling region.  For the other models the picture is qualitatively the
same\cite{be/bu}, i.e., after quantizing the theory the singularities
disappear. This has immediately the consequence that the Einstein
metric in Eq.~(\ref{08}) becomes smooth, too. Thus, the world radius in
both frames has a lower bound.

One can now ask, what is the influence of this quantization procedure
for the further evolution of the universe? For the derivation of our
results it was crucial that the solution decouples in a 2-D (dilaton
gravity) part and a 3-D spherical part. This is valid only if one
considers the theory, e.g.,inside the wormhole of fig.\ 1b. Extending
this procedure to the region away from the wormhole seems to be
difficult. But nevertheless, quantum corrections inside the wormhole
can form a dilaton potential which could be a source of an
inflationary period in later times.  A dilaton potential in our
original action (\ref{01}) or (\ref{03}) corresponds to an additional
tachyon contribution in the 2-D action (\ref{12}). The tachyon we have
discussed so far is only {\em one} possibility. Although this solution
has the advantage that the renormalization group $\beta$ functions
vanish, and thus, yielding a finite 2-D quantum field theory there are
other possible tachyon fields.  The most general tachyon field is a
combination of the solutions of the Weyl invariance condition,
which are given by
\be  \label{25}
\ba{ll}
\Phi(x,y) = a x + b y & \qquad \mbox{with} \qquad a^2 - b^2 = 
 - \kappa \qquad ,\\
T(x,y) \sim e^{\alpha x + \beta y}  & \qquad \mbox{with} \qquad
    \frac{1}{2} (\alpha^2  - \beta^2) -  a \alpha +
   b \beta - 2 = 0 \ .
\ea
\ee
In order to get the right classical limit we set furthermore $a=0$. But
there is also another parameterization for the
tachyon\footnote{For throwing our attention on this possibility we are
grateful to S. F\"orste}. Using the mass shell condition we can
replace $\alpha$ or $\beta$ and then we can expand the tachyon field
in powers of the remaining $\alpha$ or $\beta$. Since the tachyon
$\beta$ function is a linear equation in $T$ every term in this
expansion fulfills the $\beta$ equation, too. In the language of 2d
conformal field theories, this means that every term is an allowed
(1,1) operator. This procedure for constructing of new vertex
operators was described by Kawai and Nakayama\cite{kawa}. After
this procedure we find an infinite set of vertex operators or tachyon
fields which are parameterized by two integers $m$ and $n$. If we
restrict ourselves on $\kappa = \frac{24-N}{6} = 4$ (i.e.~$N=0$) these
additional terms are
\be \label{26}
T_{2}^{(n)} = (y-x)^n \, e^{2 x}  \quad , \quad T_{3}^{(m)} = (x\pm y)^m
 \, e^{2 y} \ .
\ee
Instead of Eq.~(\ref{25}) we have now as general tachyon field $T(x,y)$
\be  \label{27}
T(x,y) = \lambda e^{\frac{2}{\sqrt{\kappa}} (x-y)} + \sum_{(n,m)} 
(\mu_{2}^{n} T_{2}^{(n)} + \mu_{3}^{m} T_{3}^{(m)} )
\ee
where the function $x$ and $y$ are given by the Eq.~(\ref{22}) (the
term $T_{2}^{(0)}$ was already discussed in Refs.~5 and 14). 
A remarkable property of these terms is, that they have
in the classical limit ($\phi \rightarrow - \infty$) the typical
non-perturbative structure
\be  \label{28}
T_{2,3} \sim e^{-\frac{1}{2} e^{-2\phi}} \sim 
e^{- \frac{1}{(2 g^2_s)}}
\ee
where $g_s = e^{\phi}$ is the string coupling constant and we have
used that $h, \bar{h} \sim {\cal{O}}(e^{2 \phi})$ (because they are
quantum corrections). On the other side, in the strong coupling region
($\phi \rightarrow \infty$) we have
\be  \label{29}
T_{2} \sim e^{4 \phi} \rightarrow \infty \quad , \quad 
T_{3} \sim e^{-2 \phi} \rightarrow 0 
\ee
Therefore, these terms vanish very rapidly in the weak coupling
(classical) region and become important in the strong coupling
region. Furthermore, since $x$ and $y$ are functions of the dilaton
and moduli field\footnote{ In order to get the moduli field
$\rho$ from the Weyl field $\sigma$ one has to go from the conformal
gauge in (\ref{14}) to the Schwarzschild gauge ($ds^2 = \rho^2 dw^2 -
d\eta^2$).} \ \ these additional tachyon terms represent a
dilaton/moduli potential created by non-perturbative quantum
corrections in the strong coupling region. It remains an open question
whether this potential can yield sufficient inflation.  In order to
discuss this question one has to transform the theory back to the 4-D
Einstein frame and has to show that the resulting potential has a
flat direction which can then yield to extended
inflation\cite{ga/qu}. Probably, this is possible for a suitable
choice of the constants $\mu_{2,3}^{m,n}$. However, normally in
discussing of non-perturbative corrections one imposes further string
symmetries to restrict the possible contributions\cite{ba/di} \ and it
deserves further investigations to show that this will not destroy a
flat direction.

\section{Conclusions}
The aim of this paper was to show how quantum corrections modify a
cosmological solution of the string effective action. The classical
solution was obtained by a 5-D Kaluza-Klein approach, with an
antisymmetric tensor and dilaton field as matter part.  The 4-D FRW
metric in the string frame is completely smooth as long as the torsion
is non-vanishing. The corresponding magnetic charge prevents the
universe from collapsing. The solution is oscillating for $k=+1$ and
a time-like wormhole for $k=-1$. But nevertheless, the scalar fields,
the dilaton and the moduli, are singular at the points of minimal and
maximal extension of the universe. The moduli field or
compactification radius, e.g., is infinite inside the wormhole and
shrinks then with the time (dynamical compactification). For the
quantization of this classical theory it was crucial that near the
singularities the 5-D theory factorize in a singular 2-D part and a
smooth (spherical) 3-D part. As a first step, after a s-wave reduction
we have quantized the singular part only.  From the 4-D point of view
this partial quantization is a quasi-classical approximation, where we
quantized the scalar matter part (dilaton and moduli) only and leave
the metric and antisymmetric tensor as a classical background. As result of
this procedure the singularities in the scalar fields
disappeared and a dilaton/moduli potential can be formed. The
smoothness, especially of the dilaton fields, had the consequence that
the 4-D Einstein metric became nonsingular, too. The potential, on the
other side, was created by non-perturbative quantum corrections.  We
have discussed in principle what type of potential is possible from
the quantization of the scalar fields. But before one starts to discuss
an inflationary period driven by this potential one has first to
restrict this potential by imposing of other string symmetries which
remains a question of further investigations.

\section{Acknowledgements}
The author is grateful to Juan Garc\'{\i}a--Bellido for useful
comments. The work is supported by a DAAD grant.

\end{document}